\documentstyle[12pt]{article}

\newcommand{\eps}{\epsilon}
\newcommand{\hp}{\hat p}
\newcommand{\be}{\begin{equation}}
\newcommand{\ee}{\end{equation}}
\newcommand{\nn}{\nonumber}
\newcommand{\bea}{\begin{eqnarray}}
\newcommand{\eea}{\end{eqnarray}}

\newcommand{\Z}{Z \!\!\! Z}
\newcommand{\R}{{\kern+.25em\sf{R}\kern-.78em\sf{I} \kern+.78em\kern-.25em}}
\newcommand{\N}{{\kern+.25em\sf{N}\kern-.78em\sf{I} \kern+.78em\kern-.25em}}
\newcommand{\C}{{\kern+.25em\sf{C}\kern-.50em\sf{I} \kern+.50em\kern-.25em}}
\makeatletter
\@addtoreset{equation}{section}
\makeatother

\begin{document}

\begin{flushright}
HU-EP-02/43
\end{flushright}

\vspace*{5mm}

\begin{center}

{\LARGE\bf A Ginsparg-Wilson Approach} \\

\vspace*{8mm}

{\LARGE\bf to Lattice Chern-Simons Theory} 

\vspace*{2cm}

W. Bietenholz$^{\rm \, a}$ \ and \ P. Sodano$^{\rm \, b}$ \\

\vspace*{8mm}

$^{\rm a}$ Institut f\"{u}r Physik, Humboldt Universit\"{a}t zu Berlin \\
Newtonstr.\ 15, D-12489 Berlin, Germany \\

\vspace*{3mm}

$^{\rm b}$ Dipartimento di Fisica e Sezione I.N.F.N. \\ 
Universit\'{a} di Perugia \\ 
Via A. Pascoli, I-06123 Perugia, Italy

\end{center}

\vspace*{1cm}

{\it The concept of lattice modified symmetry formulations
is adapted to the parity symmetry of gauge fields 
and applied to the pure Abelian Chern-Simons action in three dimensions.
We derive an analogue of the Ginsparg-Wilson relation for the
parity anti-symmetry, which is motivated from the perfect lattice action,
and which we denote as the Chern-Simons-Ginsparg-Wilson relation (CSGWR).
In addition to the overlap type solutions, we construct
explicitly simple and local polynomial solutions to the CSGWR.
We show that these actions are exactly invariant under a lattice
modified parity transformation. That transformation is local as well,
and it turns into the standard parity transformation in the
continuum limit.}

\newpage

\section{Introduction}

In odd space-time dimensions, there is a possibility of adding
a gauge invariant topological Chern-Simons (CS) term to the gauge
field action. The CS term breaks both, the parity and the time-reversal
symmetry, and --- when combined with a Maxwell or Yang-Mills term ---
it leads to massive gauge excitations \cite{Jackiw}.
For an Abelian model in three space-time dimensions, the 
{\em pure CS Lagrangian} takes the form
\begin{equation} \label{CScont}
{\cal L}_{CS}[A] = \frac{\kappa}{2} A_{\mu} \epsilon_{\mu \alpha \nu}
\partial_{\alpha} A_{\nu} \ ,
\end{equation}
where $\kappa$ is a dimensionless coupling constant, and $\epsilon$
is the anti-symmetric unit tensor.

The pure CS theory is a topological field theory \cite{Witten}.
Being dominant at large distances, the CS action may be used
as a low energy effective field theory for condensed matter
systems \cite{condens}.

On the {\em lattice} the CS theory has been studied by using the Hamiltonian
formalism \cite{Semenoff}, by introducing a mixed CS action with
two gauge fields of opposite parity \cite{condens}, or by means of 
two gauge fields living on the links of two dual lattices \cite{dual}.

While in the continuum the pure CS theory is exactly soluble,
the si\-tuation is quite different on the lattice: 
the kernel
defining the naive CS action exhibits a set of zeros which are not due
to gauge invariance and the theory is not integrable even after
gauge fixing \cite{FM}. Since the Lagrangian (\ref{CScont}) is of first order
in the derivative, the appearance of extra zeros in its naive lattice
formulation is reminiscent of the doubling of fermions on the 
lattice. It has been recently shown \cite{BDS} that
the non-integrability of the CS kernel is a general feature
of any gauge invariant, local
\footnote{Throughout this paper, we define locality on the lattice
such that the couplings decay at least exponentially in the
distance, which corresponds to analyticity in momentum space.}, 
parity-odd and cubically symmetric gauge 
theory on an infinite Euclidean lattice, provided that the non-compact
link variables are parity-odd,
\begin{equation}  \label{Pstandard}
A_{\mu}(\vec x ) \to -A_{\mu} (-\vec x ) \quad
{\rm and} \quad  A_{\mu}(\vec p ) \to -A_{\mu} (-\vec p ) \ .
\end{equation}
Here $\vec x$ are the link centers and $\vec p $ are the 
corresponding momenta.
Since an additional Maxwell term regularizes the CS action
\cite{BDS}, the presence of extra zeros did
not cause problems in previous investigations of the Maxwell
CS action on the lattice.
Adding a Maxwell term to the CS action
opens a gap in the spectrum with a mechanism similar to Wilson
fermions, where a gap is opened and the unphysical poles at the
corners of the Brillouin zone are shifted to large energies.
However, in the case of lattice fermions it is possible to keep
track of the chiral symmetry much more closely than it is done
in the case of Wilson fermions, by requiring the anti-commutator
of the inverse Dirac operator with $\gamma_{5}$ to be local.
This is known as the Ginsparg-Wilson relation (GWR) \cite{GW}; it excludes 
additive mass renormalization \cite{Has} and it preserves chiral 
symmetry in an exact, though lattice modified form \cite{ML}.

Inspired by this analogy with fermionic models,
we adapt the Ginsparg-Wilson approach to CS
theories. This amounts to modifying the standard definition 
(\ref{Pstandard}) of parity on the lattice. Such an approach has been 
followed in a interesting paper by Fosco and L\'{o}pez \cite{FL};
however, their modified parity transformation is non-local,
which is problematic in view of its continu\-um limit.
\footnote{To be explicit: the lattice action in Ref.\ \cite{FL}
is local, but the lattice modified parity transformation includes
in leading order of the momentum a non-local term
$p_{\mu}p_{\nu}/p^{2}$.} 
In this paper we construct a Ginsparg-Wilson approach
to lattice CS gauge theory in which the modified parity transformation,
as well as the action itself, is manifestly {\em local}.
\footnote{A synopsis of this work was anticipated in Ref.\ \cite{Lat02}.}

For fermions in odd dimensions, a lattice modified parity symmetry
was worked out already in the framework of the GWR \cite{BN}.
The so-called parity anomaly is reproduced correctly in this way,
i.e.\ different types of 3d Ginsparg-Wilson fermions cover
all parity universality classes. 
However, in that formulation the parity transform of the
gauge field is standard, so in the case of
a pure gauge action a new transformation for
the gauge fields is needed.

The search for an integrable CS kernel on the lattice is
motivated by the attempt to compute topological invariants
in a lattice gauge theory. There are, for instance, potential 
applications to polymer physics \cite{Ferrari}. \\

In Section 2 we discuss the perfect CS lattice action, which motivates
the formulation of the Chern-Simons-Ginsparg-Wilson relation (CSGWR) as
a criterion for a lattice modified parity anti-symmetry.
In Section 3 we solve the CSGWR along the lines of the overlap formula.
However, we then concentrate on a simple polynomial solution,
which is constructed in Section 4. 
This solution does not have a fermionic analogue.
In Section 5 we derive the corresponding locally
modified parity transformation. Section 6
is devoted to our conclusion and an outlook regarding the topological
properties of such lattice formulations.

\section{A perfect Chern-Simons lattice action}

The GWR for lattice fermions was first observed
for the perfect action of free fermions, and then adapted
as a general condition for a lattice modified but exact
chirality in even dimensions 
\cite{GW,Has,ML}. For fermions in odd dimensions
it provides a lattice modified parity symmetry \cite{BN}.
In both cases, the anomalies are reproduced correctly due
to a non-trivial transformation of the fermionic measure under the
lattice modified chiral resp.\ parity transformation.

We are going to carry out the analogous step for the pure
Chern-Simons action. For comparison, we first briefly review
the perfect action for free lattice fermions. We follow the steps
presented in detail in Refs.\ \cite{qua-glu}.

\subsection{Review of the motivation for the fermionic 
Ginsparg-Wilson relation}

We consider a free, massless fermion in $d$ dimensional
Euclidean space. Its action in the continuum,
\be
s [\bar \psi , \psi ] = \frac{1}{(2\pi )^{d}} \int d^{d}p \
\bar \psi (-p) i \gamma_{\mu} p_{\mu} \psi (p) \ ,
\ee
is chirally symmetric in even dimensions, and parity
odd in odd dimensions.

We perform a block variable renormalization group
transformation (RGT) onto a lattice of unit spacing.
The blocking factor is infinite in this case, so
the lattice fermion fields $\bar \Psi , \ \Psi$ are related
to the continuum fields as
\be  \label{blocking}
\bar \Psi_{x} \sim \int_{C_{x}} d^{d}y \ \bar \psi (y) \ , \quad 
\Psi_{x} \sim \int_{C_{x}} d^{d}y \ \psi (y) \ ,
\ee
where $C_{x}$ is a unit hypercube with center $x \in \Z^{d}$.
If we use a ``$\delta$ function blocking'', then the
above relations are equations. However, we keep the 
RGT more general --- for good reasons ---
so our statement is just that the lattice variables and the
block integrals are related.

To be explicit, the perfect lattice action $S[\bar \Psi , \Psi ]$
is given by
\bea
e^{-S[\bar \Psi , \Psi ]} &=& \int D \bar \psi D \psi D \bar \eta D \eta
\exp \left\{ -s [\bar \psi , \psi ] +
\sum_{x \in \Z^{d}} \left[ \Big( \bar \Psi_{x}
- \int_{C_{x}} d^{d}d \ \bar \psi (y) \Big) \eta_{x} \right. \right. \nn \\
&& \left. \left. 
+ \bar \eta_{x} \Big(\Psi_{x}- \int_{C_{x}} d^{d}d \ \psi (y) \Big)
+ \sum_{y \in \Z^{d}} 
\bar \eta_{x} R_{xy} \eta_{y} \right] \right\} \ . 
\label{trafo}
\eea
Here $\bar \eta,\ \eta$ are auxiliary Grassmann lattice
fields. In the limit $R \to 0$ the blocking (\ref{blocking}) is 
implemented by a $\delta$ function, but we keep the term $R$ 
more general. Performing the integrals over $\bar \eta , \ \eta$
and $\bar \psi , \psi$ we arrive at the perfect lattice action
\cite{qua-glu}
\bea
S[\bar \Psi , \Psi ] &=& \frac{1}{(2\pi )^{d}} \int_{B} d^{d}p \
\bar \Psi (-p) \Delta^{-1}(p) \Psi (p) \ , \nn \\
\Delta (p) &=& \sum_{l \in \Z^{d}} \frac{\Pi ^{2} (p + 2 \pi l)}
{i \gamma_{\mu}(p_{\mu} + 2 \pi l_{\mu})} + R(p) \ , \nn \\
B &:=& ] -\pi ,\pi ]^{3} \ , \quad
 \Pi (p) := \prod_{\mu = 1}^{d} \frac{\hat p_{\mu}}{p_{\mu}} \ ,
\quad \hat p_{\mu} := 2 \sin \frac{p_{\mu}}{2} \ .
\eea

We see that the perfect action is again chirally resp.\ parity
symmetric in the usual sense if and only if $R=0$.
In this case, however, the Dirac operator $\Delta^{-1}$ is 
non-local, which avoids a contradiction with the 
Nielsen-Ninomiya No-Go theorem \cite{NN} resp.\ its counterpart 
in odd dimensions \cite{BN}.
Therefore we better insert a non-vanishing, {\em local} term $R$.
Then the lattice action becomes {\em local} \cite{UJW,qua-glu} and
the exact symmetry ($\Delta + \Delta^{\dagger}=0$)
is replaced by the GWR
\be
\Delta + \Delta^{\dagger} = 2 R \ .
\ee 

The lattice modifications of the
symmetries avoids again a contradiction with the No-Go theorem.
However, we know that this lattice formulation is related
to the exact symmetries in the continuum solely by the 
renormalization group, hence we expect the symmetry to be
still present in some way. This is confirmed by the invariance 
under a local
\footnote{The locality of the transformation depends
on the locality of the term $R$, as discussed for example in
Ref.\ \cite{Dubna}.},
lattice modified chiral \cite{ML} resp.\ 
parity \cite{BN} transformation.
We see in eq.\ (\ref{trafo}) that a non-trivial
term $\bar \eta R \eta$ breaks the symmetry,
but such a breaking in the transformation term is only 
superficial and ultimately harmless.

In fact, the GWR can be used as a criterion for a lattice modified 
but exact symmetry also in the interacting case \cite{GW,Has,BN}.
Once this relation is established, one may also
construct new solutions, which are not related to the 
perfect action any more \cite{KN}.

\subsection{A perfect lattice Chern-Simons term}

We now want to repeat the blocking from the continuum
for Abelian gauge fields, in order to construct a perfect
Chern-Simons lattice action. This is new for itself;
moreover, it will motivate the CSGWR and its (non-perfect)
solutions to be constructed in Sections 3 and 4.

For the blocking of an Abelian continuum gauge field $a_{\mu}$, 
we follow again Ref.\ \cite{qua-glu}. The blocking scheme
amounts to the integration over all straight connections
between points in adjacent hypercubes, which are separated
by the unit vector $\hat \mu$. This implies the following
relation to the lattice gauge field $A_{\mu}$,
\be
A_{\mu ,x} \sim \int_{C_{x- \hat \mu /2}} d^{d} y \
(1 + y_{\mu} - x_{\mu}) a_{\mu} (y) +
\int_{C_{x + \hat \mu /2}} d^{d} y \
(1 - y_{\mu} + x_{\mu}) a_{\mu} (y) \ ,
\ee
where $x$ is now a link center (i.e.\ $x_{\mu}$ is a half-integer,
while the remaining components of $x$ are integers).
Note that a continuum gauge transformation 
\be
a_{\mu} \to a_{\mu} + \partial_{\mu} \varphi
\ee
just implies a lattice gauge transformation
\be  \label{gaugetrafo}
A_{\mu , x} \to A_{\mu ,x} + \phi_{x+\hat \mu /2} - 
\phi_{x - \hat \mu /2} \ , \quad \phi_{x} = \int_{C_{x}} d^{d}y \
\varphi (y) \ .
\ee
In analogy to the fermionic RGT (\ref{trafo}), we construct the
perfect lattice Chern-Simons action $S_{CS}[A]$ as
\bea 
&& \hspace*{-10mm}
e^{-S_{CS}[A]} = \int Da \, DE \,
\exp \Big\{ -s[a] \nn \\
&& \hspace*{-5mm} + \sum_{x} \Big[ i E_{\mu , x}
\Big( A_{\mu ,x} - \int_{C_{x}- \hat \mu /2} d^{d} y \
(1 + y_{\mu} - x_{\mu}) a_{\mu} (y) \nn \\
&&  \hspace*{-5mm} - \int_{C_{x}+ \hat \mu /2} d^{d} y \
(1 - y_{\mu} + x_{\mu}) a_{\mu} (y) \Big)
- \frac{1}{2} \sum_{y} E_{\mu ,x} R_{\mu \nu ,xy}
E_{\nu , y} \Big] \Big\} \ , \nn \\
&& \hspace*{-5mm} = \int Da \, DE \, \exp \Big\{ -s[a] 
+ \frac{i}{(2 \pi )^{3}} \int d^{3}p \ E_{\mu}(-p) a_{\mu}(p) 
\Pi_{\mu} (p) \nn \\
&& \hspace*{-5mm} - \frac{1}{(2\pi )^{3}} \int_{B} d^{3}p \ \Big[
i E_{\mu}(-p) A_{\mu}(p) + \frac{1}{2} E_{\mu}(-p) 
R_{\mu \nu} (p) E_{\nu}(p) \Big] \Big\} \ , \nn
\eea
where
\bea
s[a] &=& \frac{\kappa}{2} \frac{1}{(2\pi )^{3}} \int d^{3}p \
a_{\mu}(-p) c_{\mu \nu}(p) a_{\nu}(p) \ , 
\quad c_{\mu \nu}(p)
:= -i \epsilon_{\mu \alpha \nu} p_{\alpha} \ ,
\nn \\
\Pi_{\mu} (p) &:=& \frac{\hat p_{\mu}}{p_{\mu}} \prod_{\nu =1}^{3}
\frac{\hat p_{\nu}}{p_{\nu}} \ .
\eea
$E_{\mu}$ is an auxiliary lattice vector field, which is
also defined on the link centers.

The classical equation of motion for the continuum gauge field
$a_{\mu}$ reads
\be
a_{\mu,cl}(p) = \frac{i}{\kappa} c^{(inv)}_{\mu \nu} (p) E_{\nu}(p)
\Pi_{\nu}(p) \ .
\ee
Note that the matrix $c_{\mu \nu}$ is anti-symmetric and therefore
singular. Hence the symbol $c^{(inv)}$ represents a {\em regularized}
inversion of $c_{\mu \nu}$, which can be obtained for instance by 
adding a tiny kinetic term to $s[a]$.  We do not specify this 
regularization here, but just fix after regularization the property
\be
c_{\mu \rho}^{\rm reg} c_{\rho \nu}^{(inv)} = \delta_{\mu \nu} \ .
\ee
Inserting the classical continuum gauge field is equivalent
to carrying out the Gaussian functional integral $\int Da$.
It leads to
\bea
e^{-S[A]} &=& \int DE \exp \Big\{ -\frac{1}{(2\pi )^{3}} \int_{B}
d^{3}p \, \Big[ i E_{\mu}(-p) A_{\mu}(p) \nn \\
&& + \frac{1}{2} E_{\mu}(-p)
C_{\mu \nu}^{(inv)}(p) E_{\nu}(p) \Big] \Big\}
\eea
and the integration $\int DE$ yields
\bea
&& \hspace*{-17mm}
S[A] = \frac{1}{(2\pi )^{3}} \int_{B} d^{3}p \, A_{\mu}(-p)
C_{\mu \nu}(p) A_{\nu}(p) \nn \\
&& \hspace*{-17mm}
C_{\mu \nu}^{(inv)}(p) = R_{\mu \nu}(p) + \frac{2}{\kappa}
\sum_{\ell \in \Z^{3}} c_{\mu \nu}^{(inv)}(p + 2 \pi \ell )
\, \Pi_{\mu}(p+2\pi l) \, \Pi_{\nu}(p+2\pi l) \ .
\eea
At this stage, the perfect lattice kernel $C_{\mu \nu}$ can be
extracted by removing the regularization incorporated in
the term $c_{\mu \nu}^{\rm reg}$.
We assume $R$ to be {\em even} with respect to standard parity,
$R_{\mu \nu}(p) = R_{\mu \nu}(-p)$, hence
\bea
C_{\mu \nu}^{(inv)}(p) + C_{\mu \nu}^{(inv)}(-p) &=& 2 R_{\mu \nu}(p) 
\qquad {\rm resp.} \nn \\
C_{\mu \nu}(p) + C_{\mu \nu}(-p) &=& 2 C_{\mu \rho}(-p) R_{\rho \sigma}(p)
C_{\sigma \nu}(p) \ ,
\eea
which we call the {\em general CSGWR}.
A finite term $R_{\mu \nu}$ breaks the parity 
anti-symmetry superficially in the lattice action $S_{CS}[A]$.\\

In particular, for the choice 
\be
R_{\mu \nu}(p) = \frac{1}{2} \delta_{\mu \nu}
\ee
we arrive at the CSGWR in the simple form that we are going to use
below,
\be  \label{CSGWR}
C_{\mu \nu} + C_{\mu \nu}^{P} = C_{\mu \rho}^{P} C_{\rho \nu} \ ,
\ee 
where $C^{P}_{\mu \nu} := P C_{\mu \nu} P$, and
$P$ is the standard parity transformation operator.

\section{Overlap-type solution to the CSGWR}

First we remark that one can solve the CSGWR in analogy 
to the fermionic overlap formula. 
To this end, it is convenient to start from the formulation
\be  \label{GamGam}
\Gamma^{P}_{\mu \rho} \Gamma_{\rho \nu} = \delta_{\mu \nu} \ , \quad
\Gamma_{\mu \rho} := C_{\mu \rho} - \delta_{\mu \rho} \ .
\ee
Let us take some
lattice CS kernel $C_{\mu \nu}^{(0)}$ to start with, such
as the Fr\"{o}hlich-Marchetti formulation \cite{FM} (its form is
given in footnote \ref{fFM}).
We assume the correct continuum limit and the absence of 
doublers for $C_{\mu \nu}^{(0)}$, but we do not require any form
of parity symmetry on the lattice. Hence the corresponding term 
$\Gamma_{\mu \nu}^{(0)} = C_{\mu \nu}^{(0)}
- \delta_{\mu \nu}$ will in general not obey relation (\ref{GamGam}),
but we can enforce the CSGWR by the transition to
\be  \label{g-ov}
\Gamma_{\mu \nu} = \Gamma^{(0)}_{\mu \rho} \cdot 
\Big( \Gamma^{(0) P} \, \Gamma^{(0)} \Big)^{-1/2}_{\rho \nu} \ ,
\ee
which is a sort of gauge overlap formula \cite{BPpriv}.
This is similar to the formula introduced in Refs. \cite{KN} for
the lattice Dirac operator $D$, where the Wilson operator $D_{W}$
was inserted as $D_{0}$; the use of kernels $D_{0}$ different
from $D_{W}$ was motivated in Refs.\ \cite{EPJC}.\\

We do not discuss further properties of this type of solution here;
instead we are going to show that simpler
and more practical solutions exist, where further properties of interest
--- gauge invariance, locality of the action and of the 
corresponding modified parity transform --- will be demonstrated.

\section{A polynomial solution to the
Chern-Simons-Ginsparg-Wilson relation}

We start from a general form of a CS lattice action,
\be
S [A] = \frac{\kappa}{2}\frac{1}{(2\pi )^3} 
\int_{B} d^{3} p \, A_{\mu}(-p) C_{\mu \nu}(p) A_{\nu}(p) \ ,
\ee
again on a 3d unit lattice in Euclidean space.
The kernel $C_{\mu \nu}$ is to be specified. For practical
reasons we put the non-compact link variables
$A_{\mu}$ on the link centers again.

Our first concern is solving the CSGWR, eq.\ (\ref{CSGWR}),
which is equivalent to eq.\ (\ref{GamGam}).
We make an explicit ansatz for the tensor $\Gamma$
in momentum space,
\bea  
&& \hspace*{-13mm}
\Gamma_{\mu \nu}(p) = \delta_{\mu \nu} u(p) 
+ L_{\mu \nu}(p) v(p) + M_{\mu \nu}(p) w(p) \ , \nn \\
&& \hspace*{-13mm}
L_{\mu \nu} (p) := - i\eps_{\mu \alpha \nu} \hp_{\alpha} \ , \quad
M_{\mu \nu}(p) := \hp^{2} \delta_{\mu \nu} - \hp_{\mu} \hp_{\nu} \ ,
\quad 
\hat p^{2} := \sum_{\alpha =1}^{3} \hat p_{\alpha}^{2} \ .
\label{Gamma}
\eea
The coefficients $u,~v,~w$ in ansatz (\ref{Gamma}) are all
assumed to be {\em parity-even}, {\em local} and 
symmetric under permutation of the axes. 
We denote the latter property as ``{\em lattice isotropy}\,''.
Moreover, the continuum limit imposes
\be  \label{cont-con}
u(p) = -1 + O(p^{2}) \ , \quad 
v(p) = \frac{\kappa}{2} + O(p^{2}) \ .
\ee
The function $w(p)$ does not have a constraint of that kind;
it must just be non-vanishing.
\footnote{As a special case, the Fr\"{o}hlich-Marchetti kernel
corresponds to $u=-1$, $v=w=\kappa /2$. \label{fFM}}

The linear term $L_{\mu \nu}$ is parity-odd and it
provides the naive discretization of the kernel $C_{\mu \nu}$.
On the other hand, the {\em Maxwell term}
$M_{\mu \nu}$ is parity-even, hence it breaks the odd parity 
symmetry of $C_{\mu \nu}$.
We know from Ref.\ \cite{BDS} that some parity-even ingredient
in $C_{\mu \nu}$ is needed in order to avoid the doubling problem.

To compute the product $\Gamma^{P} \Gamma$, we make use of
the identities
\bea
L_{\mu \rho}^{P}(p) L_{\rho \nu}(p) &=& 
[ \delta_{\mu \beta} \delta_{\alpha \nu} - \delta_{\mu \nu}
\delta_{\alpha \beta} ] \ \hp_{\alpha} \hp_{\beta} 
= - M_{\mu \nu} (p) \ ,  \nn \\
M_{\mu \rho}^{P} M_{\rho \nu} &=& 
(\hp ^{2})^{2} \delta_{\mu \nu} - \hp^{2} \, \hp_{\mu} \hp_{\nu} 
= \hp^{2} M_{\mu \nu} \ .  \label{id2}
\eea
These are simple lattice analogues to the well-known
relations in the continuum. They occur in this form
because we have assumed lattice isotropy for the terms
$v(p)$ and $w(p)$.

These two properties are crucial for the polynomial solution of 
the CSGWR, because they imply that the right-hand side reproduces
the same structure as the ansatz for $\Gamma$, up to some
coefficients (which are momentum dependent, however). 

One may compare the ansatz (\ref{Gamma}) with a Wilson fermion,
where the Maxwell term plays the r\^{o}le of the Wilson term,
which also breaks parity in odd dimensions. In both cases,
one adds an $O(a)$ suppressed parity-even term that removes
the doublers. However, the
terms in the Wilson-Dirac operator do not have any properties
analogous to the relations (\ref{id2}), 
hence in the framework of that structure one cannot find any
solution of the (fermionic) GWR. 
\footnote{If one tries to insert the Wilson Dirac operator
$D_{W}(x,y)$ into the GWR one arrives at a term $R$ which is 
non-local; for instance for the free fermion in $d=4$ ($d=2$) 
it decays only as $\vert x-y\vert^{-6}$ ($\vert x-y\vert^{-4}$)
\cite{Dubna}, which is not acceptable as a solution of the GWR.}\\

Based on the identities (\ref{id2}) we arrive at
\be
\Gamma_{\mu \rho}^{P} \Gamma_{\rho \nu} =
u^{2} \delta_{\mu \nu} + \Big[ 2 uw - v^{2} + w^{2} \hp^{2} \Big]
M_{\mu \nu} \ .
\ee
In view of condition (\ref{cont-con}), it is inevitable to set
\be  \label{ufix}
u = -1 \ .
\ee
Hence the breaking of the anti-parity of $C_{\mu \nu}$ is solely
due to the Maxwell term. For the coefficients $v,~w$
we have to require
\be
v^{2} = -2 w + w^{2} \hp^{2} \ .
\ee
This condition allows for many solutions.
It fixes, however, $w(p=0) = -\kappa^{2}/8$.
More generally, solving for $w$ and requiring locality 
\footnote{This requirement singles out one of the two 
solutions for $w(p)$ in terms of $v(p)$.}
yields
\be  \label{vw}
w = \frac{1}{\hp^{2}} \Big[ 1- \sqrt{1 + \hp^{2} v^{2}} \, \Big] \ .
\ee

At this point, we keep the maximal freedom in the 
choice of the functions $v$ and $w$, because we will need
it later on, in order to fulfill a further criterion.
We also see that the Fr\"{o}hlich-Marchetti action --- or any
other action with constant coefficients $v$ and $w$ --- fails
to solve the CSGWR.

\subsection{Gauge invariance}

We now probe the effect of a general gauge transformation
for the Abelian gauge field that we are considering. The definition
requires first another lattice discretization of a derivative,
which is actually ambiguous. However, in the spirit of the above
formulation of the action, it is obvious to choose the symmetric
nearest-neighbor formulation, as we did before in eq.\ (\ref{gaugetrafo}),
\be
A_{\mu}(p) \rightarrow A'_{\mu}(p) 
= A_{\mu}(p) - i \hp_{\mu} \phi(p) \ .
\ee
The difference between the action after and before this gauge 
transformation reads
\begin{eqnarray}
\hspace*{-8mm}
S[A'] - S[A] &=& \frac{\kappa}{2}
\frac{1}{(2\pi)^{3}} \int_{B} d^{3}p \
\Big[ i \phi (-p) \hp_{\mu} C_{\mu \nu}(p) A_{\nu}(p) \nn \\
\hspace*{-8mm}
&& -i A_{\mu}(-p) C_{\mu \nu}(p) \hp_{\nu} \phi (p)
- \phi (-p) \phi (p) \hp _{\mu} C_{\mu \nu}(p)
\hp_{\nu} \Big] .
\end{eqnarray}
Obviously, this difference vanishes if
\be \label{g-i}
\hp_{\mu} C_{\mu \nu}(p) = C_{\mu \nu}(p) \hp_{\nu} = 0 \ .
\ee
It is easy to see that
\begin{eqnarray*}
\hp_{\mu} L_{\mu \nu}(p) & = & -i \eps_{\mu \alpha \nu}
\hp _{\mu} \hp_{\alpha} = 0 = L_{\mu \nu} (p) \hp_{\nu} \ , 
\quad {\rm and} \\
\hp_{\mu} M_{\mu \nu}(p) & = & \hp_{\mu} [ \delta_{\mu \nu} \hp^{2}
- \hp_{\mu} \hp_{\nu}] = \hp^{2} \hp_{\mu} \delta_{\mu \nu} -
\hp^{2} \hp_{\nu} = 0 = M_{\mu \nu}(p) \hp_{\nu} \ .
\end{eqnarray*}
Our kernel $C_{\mu \nu}(p)$ is a
linear combination of $L_{\mu \nu}$ and $M_{\mu \nu}$, 
\be
C_{\mu \nu}(p) = L_{\mu \nu}(p) v(p) + M_{\mu \nu}(p) w(p) \ ,
\ee
where the coefficients $v(p)$ and $w(p)$ are lattice isotropic.
Therefore it obeys the property (\ref{g-i}), and we conclude
that our lattice CS action --- given by eqs.\ (\ref{Gamma}),
(\ref{ufix}) and (\ref{vw}) --- is indeed gauge invariant.

\section{A lattice modified parity transformation}

We start from a quite
general ansatz for a lattice modified parity transformation
of the gauge field,
\footnote{At this point, our notation distinguishes between the 
column vector field $A_{\mu}$ and its transpose $A_{\mu}^{t}$,
in order to avoid confusion. In the rest of this paper we simply
write $A_{\mu}$ in both cases, as it usually done.}
\begin{eqnarray}
A_{\mu}^{P} &=& P ( \delta_{\mu \nu} + X_{\mu \nu}) A_{\nu} \ , \nn \\
(A_{\mu}^{t})^{P} &=& A_{\nu}^{t} 
(\delta_{\nu \mu} + X_{\nu \mu}) P \ ,
\end{eqnarray}
where the tensor $X$ is of $O(a)$.\\

We want our lattice CS term to be {\em odd with respect
to such a modified parity transformation}. 
\footnote{We repeat that the two terms in $C = Lv + Mw$
are odd resp.\ even with respect to {\em standard} parity $P$.
However, we require the entire term $C$ to be exactly odd under 
the modified parity, which we are going to construct.}
This amounts to the requirement
\be  \label{Pmod}
C_{\mu \nu} = - [ \delta_{\mu \rho} + X_{\mu \rho}] \, C_{\rho \sigma}^{P}
\, [\delta_{\sigma \nu} + X_{\sigma \nu} ] \ .
\ee
We insert $C_{\mu \nu} = L_{\mu \nu} v + M_{\mu \nu} w$ and arrive at 
the general condition
\begin{eqnarray}
2 M_{\mu \nu} w &=& X_{\mu \rho} [L_{\rho \nu}v - M_{\rho \nu}w]
+ [L_{\mu \rho} v - M_{\mu \rho} w] X_{\rho \nu} \nn \\
&&+ X_{\mu \rho} [ L_{\rho \sigma} v
- M_{\rho \sigma}w] X_{\sigma \nu} \ . \label{conX}
\end{eqnarray}
To search for explicit solutions, we also make an
ansatz for $X$. We assume it to have the same structure as $C$,
\be
X_{\mu \nu} = L_{\mu \nu} x + M_{\mu \nu} y \ ,
\ee
the functions $x(p)$, $y(p)$ being also even and lattice isotropic.
In addition we have to arrange for them to be local, see below.
Inserting this ansatz for $X$ into condition (\ref{conX}) leads to
\begin{eqnarray}
2 M_{\mu \nu} w &=& M_{\mu \nu} \Big[ 2 v x + \hp^{2}
(2 v x y - 2 w y - w x^{2}) - (\hp^{2})^{2} w y^{2}\Big] \nn \\
&& + L_{\mu \rho} M_{\rho \nu} \Big[ 2 (vy - wx) + v x^{2} + 
\hp^{2} ( v y^{2}-2w x y) \Big] \ .
\end{eqnarray}

All in all, we have now three non-linear constraints for the
four coefficients $v$, $w$, $x$, $y$ :
\begin{eqnarray}
v^{2} + 2 w - w^{2} \hp^{2} &=& 0 \label{con1} \\
2(w-vx)(1+y \hp^{2}) + w \hp^{2}(x^{2}+\hp^{2}y^{2})
&=& 0 \label{con2} \\
2 (vy - xw) + v x^{2} + y \hp^{2}(vy-2xw)
&=& 0 \label{con3}
\end{eqnarray}
This system of equations is soluble in many ways,
but we still have to worry about locality. 

We first consider the low momentum expansion.
In agreement with eq.\ (\ref{cont-con}) we assume $v$ to take 
the form
\be
v(p) = \frac{\kappa}{2} + v_{1} \hp^{2} + O\Big((\hp^{2})^{2}\Big) \ ,
\qquad (v_{1}=const.)
\ee
which implies
\begin{eqnarray}
w(p) = - \frac{\kappa^{2}}{8} + \frac{\kappa}{2}
\Big( \frac{\kappa^{3}}{64}-v_{1}\Big) \hp^{2}
+ O\Big((\hp^{2})^{2}\Big) \ , \nn \\
x(p) = - \frac{\kappa}{4} + \frac{1}{2} \Big( 
\frac{\kappa^{3}}{64}-v_{1}\Big) 
\hp^{2} + O\Big((\hp^{2})^{2}\Big) \ , \nn \\
y(p) = \frac{\kappa^{2}}{32} + \frac{\kappa}{8} 
\Big( v_{1}- \frac{5 \kappa^{3}}{256}\Big) 
\hp^{2} + O\Big((\hp^{2})^{2}\Big) \ . \label{p-expand}
\end{eqnarray}
The existence of this expansion around $p=0$ indicates locality. 
A strict analogy to the fermionic case would suggest
$X_{\mu \nu} = -\frac{1}{2} C_{\mu \nu}$, but we see from
the expansion (\ref{p-expand}) that we have to deviate from this guess,
since it solves condition (\ref{Pmod}) only up to $O(\hp^{2})$.


Multiplying condition (\ref{con2}) and (\ref{con3})
first by $v$ resp.\ $\hp^{2}w$, and second by $w$ resp.\ $v$ ---
along with the use of condition (\ref{con1}) ---
simplifies these quadratic forms to
\begin{eqnarray}
x ( 1 + \hp^{2} y) &=& - \frac{1}{2} v \ , \nn \\
x^{2} + 2 y + \hp^{2} y^{2} &=& - w \label{con-xy} \ .
\end{eqnarray}
To solve this system explicitly, we now implement
the constraint (\ref{con1}) by inserting $v$ and $w$
as given in eqs.\ (\ref{con-xy}). If we find any solution
of the resulting equation
\be
x^{4}\hp^{2} - 2 x^{2}(1+\hp^{2}y)^{2} + \frac{1}{\hp^{2}}
\Big[ (1+ \hp^{2}y)^{4}-1 \Big] = 0 \ ,
\ee
we obtain a complete solution, with $v$ and $w$ given
by eqs.\ (\ref{con-xy}). In agreement with the low momentum
expansion and with the requirement of locality, there are 
for instance unique expressions for $v$, $w$ and $y$ in terms
of $x$,
\bea
y &=& \frac{\sqrt{1 + \hp^{2} x^{2}} -1}{\hp^{2}} \nn \\
v &=& -2x \sqrt{1 + \hp^{2} x^{2}} \nn \\
w &=& -2x^{2} \ .  \label{vwy}
\eea
This still represents the general set of local solutions.\\

To find explicit examples of solutions, we may choose
$x$ --- or any of the other coefficients --- 
so that the constraints of the low momentum expansion
is respected, and insert
it into eqs.\ (\ref{vwy}).
Then all the three constraints are guaranteed.

As a simple example we choose $x$ to be constant,
\begin{eqnarray}
x &=& -\frac{\kappa}{4} \ , \qquad
y = \frac{\sqrt{1 + \frac{\kappa^{2}}{16} \hp^{2}} -1}{\hp^{2}} \ , 
\nn \\
v &=& \frac{\kappa}{2} \sqrt{1 + \frac{\kappa^{2}}{16} \hp^{2}} \ , 
\qquad w = -\frac{\kappa^{2}}{8} \ .
\end{eqnarray}

Thus we have finally arrived at a fully explicit solution
of the CSGWR, which is exactly invariant under a lattice modified
parity transformation.
Both, the action and the transformation term, are manifestly
{\em local}.

\section{Conclusions}

We constructed a modified parity symmetry for lattice gauge
fields and applied it to the pure Abelian CS action in three
dimensions. Based on the perfect action, we derived an analogue
of the GWR for the parity anti-symmetry,
and we exhibited explicit, local solutions to this relation.
It is interesting that for gauge fields simple polynomial
solutions exist, in contrast to Ginsparg-Wilson fermions.
Unlike an earlier approach along these lines \cite{FL},
our lattice modified parity transformation is local 
as well. Hence this is a clean way to avoid the doubling
problem of the pure CS lattice action, while keeping track of the
parity anti-symmetry at finite lattice spacing.

It is well-known
that some suitable modifications of the Fr\"{o}hlich-Marchetti kernel
are integrable, still preserving locality \cite{Semenoff}.
However, those ad hoc modifications impose a constraint on the kernel
which is not compatible with the requirements of the CSGWR, and it is
not odd under a locally modified parity transformation.\\

It is an open question if the lattice CS theories
constructed in this work are topological on the lattice.
It is certain that our
polynomial solutions turn into a topological theory in the continuum
limit, but it still requires further investigation to figure out if this
property even holds on the lattice. A good check for this
property will be an explicit computation of the topological
invariant on the lattice.\\

{\bf Acknowledgment} \ {\it We are very much indebted to
Cristina Diamantini for helpful discussions, and to
Antonio Bigarini for reading the manuscript.
In addition we would like to thank 
Frederico Berruto, Philippe de Forcrand, Cesar Fosco,
Karl Jansen, Martin L\"{u}scher
and Uwe-Jens Wiese for useful comments.}


\begin{thebibliography}{50}

\bibitem{Jackiw} R.\ Jackiw and S.\ Templeton, 
{\it Phys.\ Rev.} {\bf D23} (1981) 2291.
J.\ Schonfeld, {\it Nucl.\ Phys.} {\bf B185} (1981) 157. 
S.\ Deser, R.\ Jackiw and S.\ Templeton, {\it Phys.\ Rev.\ Lett.} 
{\bf 48} (1982) 957; {\it Ann.\ Phys.} {\bf 140} (1982) 372.

\bibitem{Witten} E.\ Witten, {\it Comm.\ Math.\ Phys.} 
{\bf 121} (1989) 351;
{\it Int.\ J.\ Mod.\ Phys.} {\bf A6} (1991) 2775.

\bibitem{condens} See for example A.\ Zee, 
{\it Prog.\ Theor.\ Phys.\ Suppl.} {\bf 107} (1992) 77. 
M.C.\ Diamantini, P.\ Sodano and C.A.\ Trugenberger,
{\it Nucl.\ Phys.} {\bf B448} (1995) 585;
{\it Nucl.\ Phys.} {\bf B474} (1996) 641.

\bibitem{Semenoff} D.\ Eliezer and G.W.\ Semenoff, 
{\it Ann.\ Phys.} {\bf 217} (1992) 66;
{\it Phys.\ Lett.} {\bf B286} (1992) 118.

\bibitem{dual} R.\ Kantor and L.\ Susskind, 
{\it Nucl.\ Phys.} {\bf B366} (1991) 533.

\bibitem{FM} J.\ Fr\"{o}hlich and P.A.\ Marchetti, 
{\it Comm.\ Math.\ Phys.} {\bf 121} (1989) 177.

\bibitem{BDS} F.\ Berruto, M.C.\ Diamantini and P.\ Sodano,
{\it Phys.\ Lett.} {\bf B487} (2000) 366.

\bibitem{GW} P.H.\ Ginsparg and K.G.\ Wilson, 
{\it Phys.\ Rev.} {\bf D25} (1982) 2649.

\bibitem{Has} P.\ Hasenfratz, V.\ Laliena and F.\ Niedermayer,
{\it Phys.\ Lett.} {\bf B427} (1998) 125.
P.\ Hasenfratz, {\it Nucl.\ Phys.} {\bf B525} (1998) 401.

\bibitem{ML} M.\ L\"{u}scher, {\it Phys.\ Lett.} {\bf B428} (1998) 342.

\bibitem{FL} C.D.\ Fosco and A.\ L\'{o}pez,
{\it Phys.\ Rev.} {\bf D64} (2001) 025017.

\bibitem{Lat02} W.\ Bietenholz, J.\ Nishimura and P.\ Sodano,
{\tt hep-lat/0207010}, to appear in Nucl. Phys. B (Proc. Suppl.).

\bibitem{BN} W.\ Bietenholz and J.\ Nishimura,
{\it JHEP} {\bf 0007} (2001) 015.

\bibitem{Ferrari} F.\ Ferrari and I.\ Lazzizzera, 
{\it Nucl.\ Phys.} {\bf B559} (1999) 673.
F.\ Ferrari, {\it Annalen Phys.} {\bf 11} (2002) 255;
{\it J.\ Phys.} {\bf A36} (2003) 5083. 

\bibitem{qua-glu} W.\ Bietenholz and U.-J.\ Wiese,
{\it Phys.\ Lett.} {\bf B378} (1996) 222; 
{\it Nucl.\ Phys.} {\bf B464} (1996) 319.

\bibitem{NN} H.B.\ Nielsen and M.\ Ninomyia, 
{\it Nucl.\ Phys.} {\bf B185} (1981) 20.

\bibitem{Dubna} W.\ Bietenholz, {\it in} ``Lattice Fermions and
the Structure of the Vacuum'', Kluwer (2000),
V.\ Mitrjushkin and G. Schierholz eds., p.\ 77 
({\tt hep-lat/0001001}). 

\bibitem{UJW} U.-J.\ Wiese, {\it Phys.\ Lett.} {\bf B315} (1993) 417.

\bibitem{KN} Y.\ Kikukawa and H.\ Neuberger,
{\it Nucl.\ Phys.} {\bf B513} (1998) 735.
H.\ Neuberger, {\it Phys.\ Lett.} {\bf B417} (1998) 141;
{\it Phys.\ Lett.} {\bf B427} (1998) 353.

\bibitem{BPpriv} F.\ Berruto and P.\ Sodano, private notes.

\bibitem{EPJC} W.\ Bietenholz, {\it Eur.\ Phys.\ J.} {\bf C6} 
(1999) 537; {\it Nucl.\ Phys.} {\bf B644} (2002) 223.
W.\ Bietenholz and I.\ Hip, {\it Nucl.\ Phys.} {\bf B570} (2000) 423.
P.\ Hasenfratz, S.\ Hauswirth, K.\ Holland, T.\ J\"{o}rg and
F.\ Niedermayer, {\it Nucl.\ Phys.} {\bf B643} (2002) 280.

\end{thebibliography}
\end{document}